\documentclass[runningheads]{llncs}
\usepackage[T1]{fontenc}
\usepackage{graphicx}
\usepackage{color}
\usepackage{placeins}
\usepackage{float}
\usepackage{cleveref}
\usepackage{ulem}
\usepackage[table,xcdraw]{xcolor}
\usepackage{array}
\usepackage{textcomp}
\usepackage{subcaption}
\usepackage{adjustbox}
\usepackage{booktabs,array}

\begin{document}

\title{A Non-Intrusive Framework for Deferred Integration of Cloud Patterns in Energy-Efficient Data-Sharing Pipelines}
\titlerunning{Deferred Integration of Cloud Design Patterns}
%

\author{
Sepideh Masoudi\inst{1} \and
Mark Edward Michael Daly\inst{1} \and
Jannis Kiesel\inst{1} \and
Stefan Tai\inst{1}
}
\authorrunning{S. Masoudi et al.}
%
\institute{
\textsuperscript{1} Information Systems Engineering, Technische Universität Berlin, Berlin, Germany \\
\email{\{smi,jaki,st\}@ise.tu-berlin.de}\\
\email{m.daly.higham@campus.tu-berlin.de}\\
\url{https://www.tu.berlin/en/ise}\\
}
\maketitle              
\begin{abstract}

As data mesh architectures gain traction in federated environments, organizations are increasingly building consumer-specific data-sharing pipelines using modular, cloud-native transformation services. Prior work has shown that structuring these pipelines with reusable transformation stages enhances both scalability and energy efficiency.  However, integrating traditional cloud design patterns into such pipelines poses a challenge: predefining and embedding patterns can compromise modularity, reduce reusability, and conflict with the pipelines' dynamic, consumer-driven nature.
To address this, we introduce a Kubernetes-based tool that enables the deferred and non-intrusive application of selected cloud design patterns without requiring changes to service source code. The tool supports automated pattern injection and collects energy consumption metrics, allowing developers to make energy-aware decisions while preserving the flexible, composable structure of reusable data-sharing pipelines.

\end{abstract}
\section{Introduction}With the big data market expected to surpass \textdollar{}800 billion by 2030, more organizations are expected to adopt data-sharing practices in federated environments~\cite{grandview_big_data_2030}. One emerging approach to support this trend is the Data Mesh, a decentralized data architecture structured around domain-specific ownership. It enables data sharing, managing, and value generation from analytical data within and across enterprise boundaries~\cite{van2024architectural,datameshprinciple}.

Within a Data Mesh, data-sharing pipelines have become an established mechanism for exchanging data among independent organizations~\cite{munappy2020data}. These pipelines consist of sequential, interconnected transformation stages that adapt data to meet collaboratively defined requirements such as privacy constraints, analytical needs, or compliance policies~\cite{caise-paper}.

By allowing domain teams to implement modular, service-oriented transformation stages, Data Mesh promotes reuse: when multiple data consumers have similar requirements, stages and their outputs can be reused, reducing both effort and energy consumption~\cite{caise-paper}. 
However, designing one-to-many data-sharing pipelines introduces new complexities. The dynamic nature of consumer requirements calls for tools that support informed design decisions, enabling teams to maximize stage reuse while improving for energy efficiency.~\cite{caise-paper,falconi2024data}.

Leveraging cloud-based services for transformation stages aligns with the Data Mesh principle of decentralizing data architecture, shifting away from monolithic systems like traditional data lakes toward an interoperable, service-oriented environment~\cite{wider2023decentralized,ashraf2023key}.
In this model, data-sharing pipelines are implemented as compositions of modular, cloud-native services. These services are decoupled, reusable, independently scalable, and can be dynamically reconfigured and distributed across multiple organizational boundaries~\cite{wu2015service}.

To ensure reliability and operational efficiency, the design and configuration of these pipelines should incorporate established cloud design patterns~\cite{microsoft_azure,fehling2014cloud}.
However, applying design patterns in the context of stage-based, consumer-specific data sharing pipelines introduces new challenges. In such pipelines, the sequence and combination of transformation services can vary significantly depending on the requirements of each data consumer.
As a result, embedding cloud design patterns at the time of service implementation undermines modularity and reusability~\cite{microsoft_azure}. Instead, the appropriate pattern should be applied declaratively and deferentially, based on the specific pipeline configuration and the context in which a transformation service is used. This approach supports both reuse and dynamic adaptability across diverse data-sharing scenarios.

Given these challenges, there is a clear need for mechanisms that enable the automatic, non-intrusive application of cloud design patterns to transformation services without requiring changes to their source code. This ensures the continued reusability of modular services across pipelines with varying configurations. At the same time, there is a need to monitor the impact of applied patterns on energy consumption, enabling developers to make informed, 
energy-aware decisions during pipeline design.

This leads us to the central research question:
\textit{"How can cloud design patterns be applied in the design of data-sharing pipelines while preserving pipeline stage reusability and improving energy efficiency?"}

To address this question, we present an open-source Kubernetes-based tool that supports the deferred and non-intrusive application of selected cloud design patterns in data-sharing pipelines. The tool works by automatically deploying lightweight proxy services and updating service configurations to implement the chosen patterns. Additionally, it collects performance and energy consumption metrics, allowing developers to evaluate the effectiveness of individual pattern applications within specific pipeline contexts.

The remainder of this paper is organized as follows: In the next section, we review related work, focusing on data-sharing pipeline design and the impact of design patterns in cloud-native applications. Section 3 introduces our proposed tool for applying cloud design patterns and analyzing their impact on resource consumption. Section 4 presents the proof of concept. Finally, Section 5 concludes the paper and outlines directions for future work.
\label{introduction}
\section{Related Works}TOSCAData, proposed by Dehury et al., adopts a data pipeline as code model with reusable blocks across multiple cloud providers, yet it does not address supporting the use of cloud design patterns or offering feedback on their impact on resource consumption~\cite{dehury2022toscadata}. Likewise, Sildatke et al. introduce a distributed microservice architecture for information extraction pipelines that handles diverse data formats but does not address dynamic pattern application~\cite{sildatke2023distributed}.

Studies such as Dynamos by Stutterheim et al. demonstrate policy-driven, event‑based composition of microservices into data exchange platforms, but they do not tackle cloud design pattern concerns during service chaining~\cite{stutterheim2024dynamos}. 
In the manufacturing domain, Wu et al. encapsulate feature-based data exchange as on‑demand web services for CAD systems, showcasing scalable, service‑oriented pipelines without explicit support for design pattern selection~\cite{wu2015service}. 

In the broader cloud‑native field, several studies offer decision support and performance evaluation for design patterns. Waseem et al. map application requirements (e.g., service discovery, security) to microservice patterns via decision models~\cite{waseem2022decision}
Pinciroli et al. develop performance models to quantify the impact of selected patterns~\cite{pinciroli2023performance}, and Noureddine et al. experimentally measure the energy costs of Gang‑of‑Four patterns across platforms~\cite{noureddine2025investigating}. 
Werner et al. provide a multilayer energy measurement framework for cloud‑native applications~\cite{werner2025comprehensive}, and Xiao et al. examine energy performance maintainability trade‑offs in microservices~\cite{xiao2025effectiveness}. Zdun et al. automate conformance checks for microservice decomposition patterns~\cite{10.1007/978-3-319-69035-3_29}.
While these contributions richly inform pattern selection and evaluation, they focus on static, code‑level solutions and lack mechanisms for deferred, code-agnostic application of cloud design patterns within the dynamic, consumer-driven context of data‑sharing pipelines.
It is worth mentioning that, unlike software design patterns~\cite{gamma1995design}, which focus on code-level solutions and object interactions, cloud design patterns address deployment and scalability challenges in cloud environments~\cite{10.1145/2742854.2747280,fehling2014cloud}. While the use of software design patterns, such as Chain of Responsibility and Strategy, remains important and should be considered in the implementation of transformation services, our work focuses on cloud design patterns that can be applied independently of the source code and may differ from one configuration to another in cloud-based data-sharing pipelines.

Building on these works, we propose a tool that automates the injection of cloud design patterns into transformation services of data-sharing pipelines without altering service source code and integrates this with performance and energy consumption reporting to support pipeline specific design decisions.

\section{Non-Intrusive Deferred Integration of Cloud Patterns}Data-sharing pipelines consist of a chain of sequential stages, aiming to transform data from a data product~\cite{datameshprinciple} to meet consumer-specific requirements before exposing it to consumers. 
Filtering, aggregation, anonymization, and formatting are examples of operations performed on data in a transformation stage~\cite{caise-paper}.

The domain team within a Data Mesh must build a new and unique combination of transformation stages, modeled as a pipeline, for each consumer.
Through parameterized implementation of transformation stages, they can be reused across different data-sharing pipelines originating from the same data provider~\cite{van2024architectural,caise-paper}. However, the pipeline structure, configuration and ordering of the stages vary from one pipeline to another, depending on the specific requirements of each consumer, the data volume and the frequency of data transfer.
Therefore, appropriate design decisions differ from one pipeline to another, even when the same transformation stages are employed.

Due to the decentralized nature of data sharing pipelines in cloud environments and the need for cloud system qualities of scalability, reliability, there is a motivation to adopt cloud design patterns during the design and implementation of such pipelines.
However, the decision of which cloud design patterns to apply, at which stages, and how they would affect energy consumption should be shifted from the implementation of the transformation service to the design and configuration of each pipeline.

Thus, by decoupling the implementation of cloud design patterns from the code and the implementation of individual stages, transformation stages can be reused across pipeline structures and configurations with varying designs. Moreover, design patterns can be selected based on their impact on qualities and with energy-efficiency under constant consideration.

Although several studies aim to automate and facilitate the chaining of transformation stages into pipelines~\cite{sildatke2023distributed,stutterheim2024dynamos}, the question of how to apply cloud design patterns to these transformation stages without compromising their reusability across different pipelines and energy efficiency remains open. 
There is also a lack of support for making design decisions, based on the specific combination of transformation services, rather than relying on static, up-front implementations.

\subsection{Deferred Automatic Application of Cloud Design Patterns}
We present \textit{SnapPattern}, an open-source tool (available on GitHub\footnote{\url{https://github.com/Sepide-Masoudi/SnapPattern}}), implemented in Java, that enables the automatic, non-intrusive application of selected cloud design patterns to cloud-based applications without requiring any modifications to the service source code. SnapPattern works by dynamically updating the configurations of transformation services and deploying the auxiliary components necessary to implement the chosen patterns and monitoring tools including Kepler, OpenTelemetry, and Prometheus to support energy-aware and performance-conscious design decisions.
This approach allows developers to leverage the benefits of cloud design patterns, such as improved resilience, observability, and scalability, while preserving the modularity and reusability of transformation services across diverse pipeline configurations.

\begin{figure}
\centering
\includegraphics[width=0.90\textwidth]{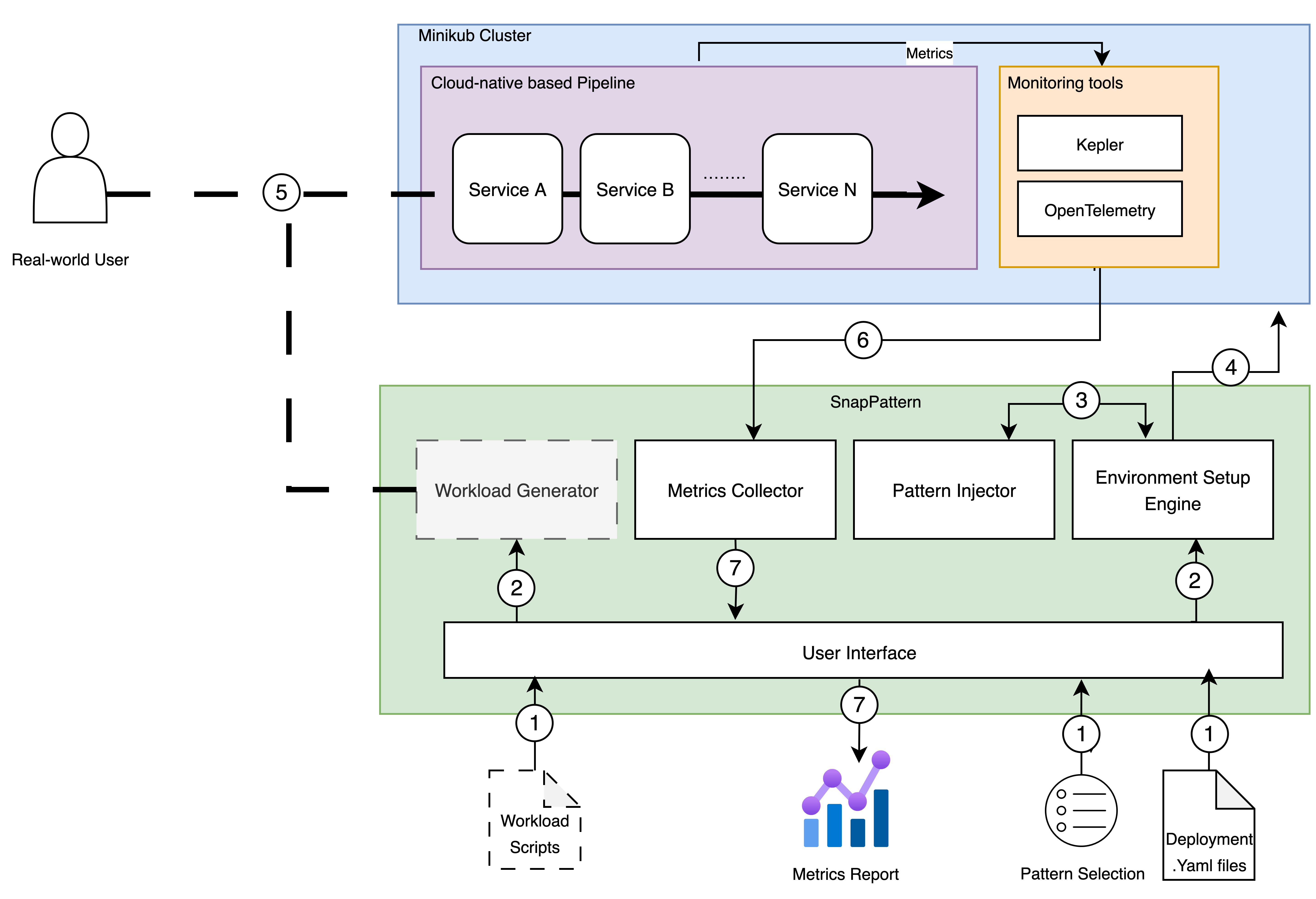}
\caption{Logical architecture of SnapPattern for the automated application of cloud design patterns.} \label{arch}
\end{figure}

As shown in \cref{arch}, the tool consists of five main components: the \textit{User Interface}, \textit{Environment Setup Engine}, \textit{Pattern Injector}, \textit{Workload Generator}, and \textit{Metrics Collector}. As the first step in using the tool, the user needs to set up the Minikube environment, which is automatically done after clicking a button by the \textit{Environment Setup Engine}. This component also supports deleting the cluster and listing the services deployed on it, filterable by namespace.
As the next step, the user needs to upload the deployment YAML file in Kubernetes format for the transformation services, which will later be deployed on the already created cluster automatically by clicking the corresponding button.
Later, the user can select the cloud design patterns to be applied, as well as the services to which the patterns should be applied. All of these configurations can be set up in a user-friendly manner through the \textit{Pattern Injector} component, just as in the previous steps.
All other tools or services needed for the application of selected cloud design patterns will also be automatically deployed in that environment through a collaboration between \textit{Pattern Injector} and \textit{Environment Setup Engine} components. 
\textit{Pattern Injector} will modify the YAML file of the microservices (if needed) to add the necessary configuration to apply the cloud design patterns and redeploy them. This component will also deploy external services and proxies that are responsible for implementing and deploying the selected design pattern outside the transformation services.

Workload scripts can be uploaded to generate synthetic workloads, handled by \textit{Workload Generator} component , or the user can skip this step and use real-world requests triggered by users. 
Finally, the \textit{Metrics Collector} will deploy the monitoring tools and export reports and plots that reveal the energy and performance related metrics. To provide insight into the exact resource consumption of the additional pattern-related services separately from the baseline pipeline services, we deploy these services in different namespaces. 
This separation allows us to calculate the total consumption of resources of both the pipeline services and the pattern services independently by the \textit{Metrics Collector} component.
At the current stage, \textit{SnapPattern} supports five cloud design patterns: \textit{Circuit Breaker (CB)}, \textit{Cache Aside (CA)}, \textit{Request Collapsing (RC)}, \textit{Gateway Offloading (GO)}, and \textit{Asynchronous Request-Reply (ARR)}~\cite{microsoft_cloud_design_patterns_2022}. 
However, the tool is extendable and it is possible to add more patterns simply by implementing the \textit{PatternGenerator} interface and providing the required resources for deploying the new patterns. We chose these five cloud design patterns because they are relevant to industry practitioners~\cite{vale2022designing} and have been indicated as relevant to performance~\cite{microsoft_cloud_design_patterns_2022}, addressing common challenges such as routing, scalability, and efficient client responses. In addition, feasibility, ease of implementation, and compatibility with external reusable implementation approaches were also considered when selecting this subset of patterns.
\label{architecture}
\section{Proof of Concepts}The proposed tool represents an early-stage effort to decouple the application of cloud design patterns from the source code of transformation services in data-sharing pipelines. Its goal is to improve the reusability of transformation services across different configurations and use cases, a necessity driven by the dynamic and consumer-driven nature of data-sharing pipelines, while also assisting in selecting which design patterns to apply and where to apply them based on pipeline requirements and configurations.
This section aims to explain the technical feasibility of the tool in more detail and to demonstrate its integration within a data-sharing pipeline scenario.
\subsection{Technical Feasibility Demonstration}
We selected a suite of production‑grade, open‑source, container‑ready tools, widely adopted in cloud environments, to implement external design patterns without modifying application code. All design patterns are implemented as non‑intrusive, Kubernetes‑managed proxy or sidecar services packaged as Docker images and configured via environment variables, ConfigMaps, and templated YAML. The Asynchronous Request Reply pattern uses a Flask‑based HTTP proxy that wraps selected paths into RabbitMQ queues and stores results in Redis, all deployed via Helm and kubectl in Minikube. Gateway Offloading relies on the NGINX Ingress Controller(installed with Helm) and advanced annotations for rate limiting and body‑size enforcement, with ingress rules generated at runtime by the tool. The Cache‑Aside pattern is implemented in two transparent, Kubernetes‑orchestrated variants without any application‑code changes: the SQL‑level approach uses ProxySQL in place of the original database DNS name (with the real backend renamed), defining regex‑based query rules, TTLs, and performance parameters(threads, connections, cache size) in a ConfigMap (proxysql.cnf) for in‑memory result caching and live tuning; the HTTP‑level alternative employs a FastAPI proxy paired with a Redis cluster to cache GET requests between services. Request Collapsing is realized through an NGINX proxy extended with Lua scripting and a shared memory dictionary, dynamically routed via a ConfigMap. Finally, the Circuit Breaker and Retry pattern runs Envoy Proxy with circuit‑breaking and retry policies defined in a generated YAML ConfigMap, mounted into the proxy container and exposed through a Kubernetes Service.

Using the GUI, a typical pattern injection workflow begins with the user uploading a Kubernetes deployment YAML file for the baseline pipeline, after which a Minikube cluster is automatically created. The user then selects a design pattern, configures its parameters(where to apply the pattern), and the tool generates and deploys the pattern to the cluster. Finally, the system monitors service readiness and confirms successful injection, enabling repeatable metrics evaluation with minimal user effort.
During each test run, Prometheus retrieves container-level metrics from Kepler and service-level telemetry from OpenTelemetry spans. The collected metrics are exported as CSV and organized into structured Excel tables for multidimensional analysis. The tool also includes Python scripts to generate illustrative plots of energy metrics.
It is worth noting that the deployment of all required infrastructure components, such as the Minikube cluster, monitoring stack, and pattern stack, is fully automated and can be initiated with a single click via dedicated buttons in the GUI, same as the deleting the cluster or design patterns.
\subsection{Functionality and Integration} 
To demonstrate the use of our tool in a representative scenario, we implemented four transformation services including filtering, aggregation, anonymization, and formatting using the Java programming language and the Spring Boot framework. 
The anonymization service supports two strategies: masking and hashing. The formatting service supports output in either CSV or JSON.
It is important to note that our tool is language-agnostic; it only requires the user to provide the corresponding Kubernetes deployment YAML files for transformation services.
We defined multiple chains of these transformation services to construct different data-sharing pipelines. 
For our data product, we used publicly available data from open data portal, which contains a list of banned books\footnote{\url{https://www.berlin.de/verbannte-buecher/suche/index.php/index/all.json?q=}}. This data product is made available to the transformation services via a dedicated service called \textit{data-product-service}.
The primary objectives of our experiments are, first, to demonstrate the usage of the proposed tool and its capability to inject cloud design patterns on-demand, and second, to show the effects of applying these design patterns by utilizing the monitoring dashboard of our tool.
To achieve this, we used our tool to deploy our mentioned pipelines. 
We applied three workload levels: low, medium, and high, where the number of user requests increases every 30 seconds, by 10, 20, and 40 users, respectively. 
Each workload was run on a minikube node with 8 CPUs and 24 GB of memory, on a Linux host equipped with an Intel x86\_64 12-core i7-1355U processor and 32 GB of memory. 

We used our tool to apply these five design patterns, one pattern at a time, to the selected services of our data-sharing pipelines and monitored their effects on the collected metrics.
Based on the logic of the system under test, the Circuit Breaker and Asynchronous Request-Reply patterns were applied to the main transformation services (Filtering, Aggregation, Anonymization, Formatting), the Gateway Offloading pattern was applied to the Coordinator Service, which is responsible for coordinating the pipelines, and the Request Collapsing and Cache Aside patterns were applied to the Data Product service which acts as a data product accessible through a REST API.

\begin{figure}[ht]
    \centering
    \includegraphics[width=\linewidth]{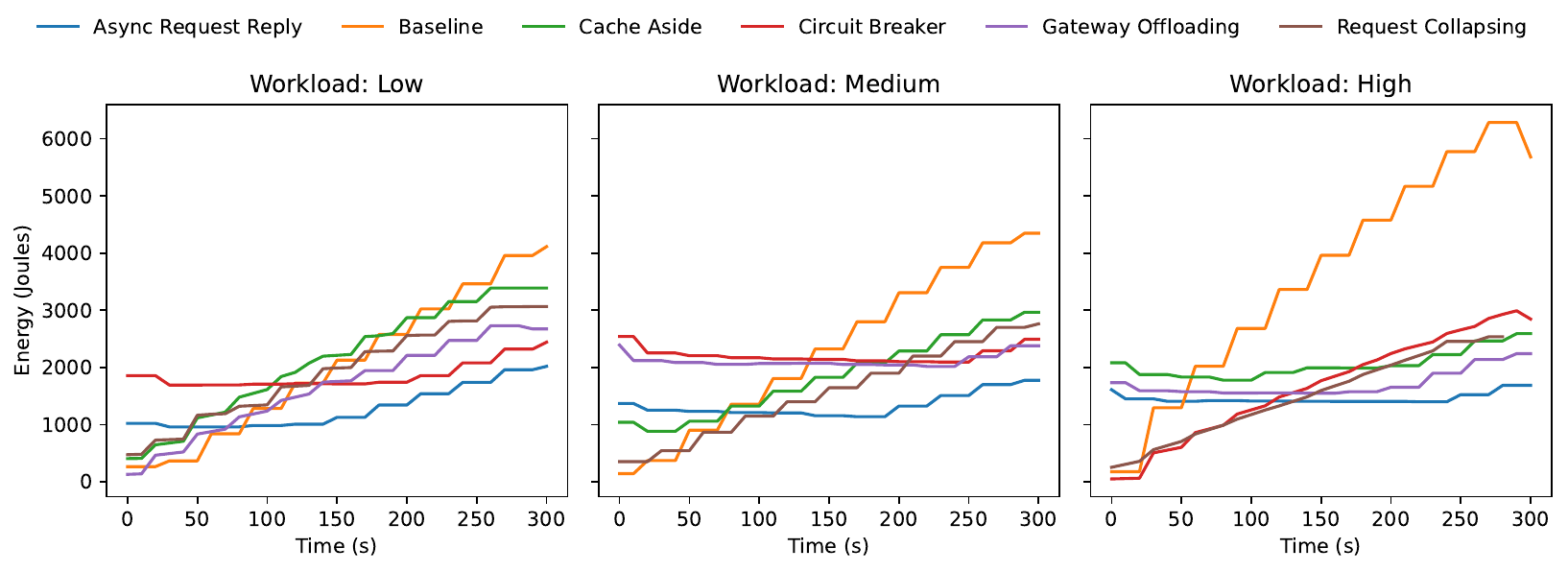}
    \caption{Energy consumption measured in joules per 10 seconds for the workload scenarios Low, Medium, and High for over all system under the test.}
    \label{fig:EnergyMeasurements}
\end{figure}

Fig. \ref{fig:EnergyMeasurements} illustrates the measured energy consumption metrics in joules per second for all pipelines. The figure shows how the energy usage of our sample data-sharing pipeline changes when used as a baseline and when automatically applying each of the design patterns using the tool, as described above. The scripts to generate these plots are available as part of the Monitoring component of the tool.
The data-sharing pipeline source code, used as the system under test, can be found on GitHub\footnote{Pipeline:\url{https://github.com/Sepide-Masoudi/Data-sharing-pipeline}}.
A video tutorial demonstrating the tool's usage in an experiment is also available in the project's GitHub repository\footnote{\url{https://vimeo.com/1100669519}}.
The experiments are easily reproducible, and the tool can be seamlessly used with any other system.
\label{evaluation}
\section{Conclusion}In this paper, we addressed the challenge of dynamically designing data-sharing pipelines within decentralized Data Mesh environments by leveraging cloud design patterns in a flexible and reusable manner. We highlighted the importance of preserving the modularity and reusability of cloud-based transformation services across pipelines with varying consumer-specific requirements.
To overcome the limitations of static, code-level pattern integration, we introduced SnapPattern, a Kubernetes-based tool that enables the automated, non-intrusive application of cloud design patterns, without modifying the source code of services. By integrating monitoring components such as Kepler, OpenTelemetry, and Prometheus, SnapPattern also provides insights into the performance and energy implications of pattern choices, energy-aware design decisions.
Future work should aim to incorporate additional design patterns, support a wider range of benchmarking tools and integrate the tool with other data platforms and workflow orchestrators, such as Argo Workflows and use the tool in real-word pipelines.


\label{conclusion}

\begin{credits}
\subsubsection{\ackname} Funded by the European Union (TEADAL, 101070186). Views and opinions expressed are however those of the authors only and do not necessarily reflect those of the European Union. Neither the European Union nor the granting authority can be held responsible for them.

\end{credits}
%
%
%

%
\end{document}